\documentclass[a4paper,11pt]{article}
\pdfoutput=1 

\usepackage{contribution}

\newcommand{\weblink}[2][]{%
    \ifthenelse{\equal{#1}{}}%
    {\textnormal{\url{#2}}}%
    {\textnormal{\href{#2}{#1}}}%
}

\def\beq{\begin{equation}}
\def\eeq#1{\label{#1}\end{equation}}
\def\eeqn{\end{equation}}

\def\beqa{\begin{eqnarray}}
\def\eeqa#1{\label{#1}\end{eqnarray}}
\def\eeqan{\end{eqnarray}}

\let\bar=\overbar

\def\Dslash{\not{\hbox{\kern-4pt $D$}}}
\def\dslash{\not{\hbox{\kern-2pt $\del$}}}

\def\msb{{\bar{\ssstyle M \kern -1pt S}}}

\newcommand{\contribution}[7][]{%
  \clearpage
  \thispagestyle{plain}
  \ifthenelse{\equal{#1}{}}
  {\hypersetup{pdftitle={#2}}}
  {\hypersetup{pdftitle={#1}}}
  \hypersetup{pdfauthor={{#3} {#4}}}
  {\centering\normalfont\LARGE\bfseries\sffamily #2 \par\nobreak}
  \lhead{}
  \chead{%
    \textit{\footnotesize XIV International Conference on Hadron Spectroscopy
      (\weblink[\textit{hadron2011}]{http://www.hadron2011.de}), 13-17 June 2011, Munich, Germany}%
  }
  \rhead{}
  \bigskip
  \begin{center}
    {#3} {#4}\ifthenelse{\equal{#6}{}}{}{\footnote{\weblink[#6]{mailto:#6}}}
    \ifthenelse{\equal{#7}{}}{}{#7} \\
    \textit{#5}
  \end{center}
  \bigskip
}

\renewcommand{\abstract}[1]{%
  \begin{center}
    \begin{minipage}{0.85\textwidth}
      \begin{footnotesize}
        #1
      \end{footnotesize}
    \end{minipage}
  \end{center}
  \bigskip
}

\begin{document}

%
%
%
%
%

{  

\newcommand{\du}{{\Delta u}}
\newcommand{\dd}{{\Delta d}}
\newcommand{\ds}{{\Delta s}}

\newcommand{\dub}{{\Delta \bar u}}
\newcommand{\ddb}{{\Delta \bar d}}
\newcommand{\dsb}{{\Delta \bar s}}

\newcommand{\dS}{{\Delta \Sigma}}
\newcommand{\dg}{{\Delta g}}
\newcommand{\dG}{{\Delta G}}

\newcommand{\ua}{{\uparrow}}
\newcommand{\da}{{\downarrow}}

%

\contribution[]  
{Nucleon Spin Structure and Parton Distribution Functions}  
{J\"org}{Pretz}  
{Physikalisches Institut  \\
  Universit\"at Bonn \\
  D-53115 Bonn, GERMANY}  
{jorg.pretz@cern.ch}  
{on behalf of the {\sl COMPASS} Collaboration}  
%

\abstract{%
This article gives an overview over recent results on quark and gluon helicity 
distributions obtained in deep inelastic lepton nucleon
scattering and proton proton interactions.
Future experimental programs to study the nucleon structure will be discussed
as well.
}
%

\section{Introduction}
In the Quark Parton Model, the nucleon is successfully described 
in terms of parton distribution functions (PDFs).
Whereas unpolarized parton distributions like $q(x)$ and $g(x)$
interpreted as number densities of quarks and gluons at a
given longitudinal momentum fraction $x$ in the nucleon
are relatively well known,
distributions involving polarization degrees of freedom
are less well known.
The most prominent ones, related to the nucleon spin problem, are the helicity
distributions $\Delta q(x)$, $\Delta g(x)$ and the
transversity distributions $\Delta q_T(x)$.

This paper focuses on the helicity distributions.
Effects occurring when considering transverse momenta and transverse polarizations
are discussed in the contribution of M. Anselmino~\cite{anselmino}.
Section~\ref{nuc_spin} gives a short summary of the nucleon spin puzzle
and its connection to helicity distributions of quarks and gluons.
Section~\ref{method} discusses various experimental methods to access the helicity
distributions. Recent results are presented in Section~\ref{results}.
Future experimental programs are discussed in Section~\ref{future}.

\section{The nucleon spin puzzle}\label{nuc_spin}
The spin of the nucleon can be decomposed in helicity ($\dS$ \& $\dG$) and orbital
angular momentum contributions ($L_q$ \& $L_g$) of quarks and gluons
\[
  \frac{1}{2} = \frac{1}{2} \Delta \Sigma + \Delta G + L_q + L_g \; .
\]
For a recent discussion on ambiguities in this decomposition see~\cite{leader1}.
Whereas the static quark model predicts $\dS=1$ and zero contribution
from $\Delta G$, $L_q$ and $L_g$, relativistic quark models predict
a helicity contribution of quarks, $\dS$ of the order of $60\%$~\cite{bass}.
Results from polarized deep inelastic scattering indicate a much smaller
value: $\dS \approx 25\%$,
allowing thus for large contributions of $\Delta G$, $L_q$ and  $L_g$. 
In recent years mainly the measurement of $\dG$ was in the focus of research
because a large contribution $\dG \approx 2-3$ could explain the 
small value of $\dS$ measured in deep inelastic scattering via the
mechanism of axial anomaly~\cite{anomaly}.
While a contribution of 400 -- 600 \% (corresponding to $\dG = 2-3$) of the gluon helicity to 
the nucleon spin may sound strange, one should keep in mind
that although perturbative QCD is not able to predict $\Delta G$,
it can predict its scale dependence.
It is given by $\dG(\mu) \alpha_s(\mu) = \mbox{const.}$
in next-to-leading order (NLO) QCD, i.e. with increasing scale $\mu$, $\dG$ must increase because
the strong coupling constant $\alpha_s$ decreases.

The helicity contributions of quarks, $\dS$ can be further decomposed 
in the contributions of different quark flavors depending on the momentum fraction $x$
carried by the quarks:
\[
   \Delta \Sigma = \int_0^1 \sum_{q} \Delta q(x) \rm{d}x
\]
where the sum runs over all light quark flavors $q= u, d, s, \bar u, \bar d, \bar s$.
The helicity distribution is defined as $\Delta q(x) =
q^\uparrow(x) - q^\downarrow(x)$, the unpolarized distributions are given by $q(x)=q^\uparrow(x) + q^\downarrow(x)$.
Similar to the unpolarized quark distributions $q^\uparrow(x)
(q^\downarrow(x))$
are number densities of quarks with spin parallel (anti-parallel) to the
nucleon spin. 

In a similar way the first moment of the gluon distribution is
given by the integral over the gluon helicity distribution
\[
  \Delta G = \int_0^1 \Delta g(x) {\rm d}x \, .
\]

As mentioned above, presently we know that $\dS$ is of the order of 25\%.
Open questions addressed in this article are the distribution of these 25\% among
the different quark flavors and new results on the gluon helicity distribution $\Delta g(x)$.

\section{Accessing the helicity distributions}\label{method}
Helicity distributions can be accessed in deep inelastic scattering and
proton-proton scattering.
The most simple case of inclusive polarized deep inelastic scattering 
\[
  \vec \ell  + \vec N \rightarrow \ell' + X
\]
will be discussed
in more detail. In all reactions mentioned $X$ stands for unobserved
final state particles. A polarized lepton creates a polarized photon.
By changing the relative spin orientation of lepton  
and nucleon the two situations shown in Fig.~\ref{spineff} are
obtained.
\begin{figure}
\centering{\includegraphics[width=0.65\textwidth]{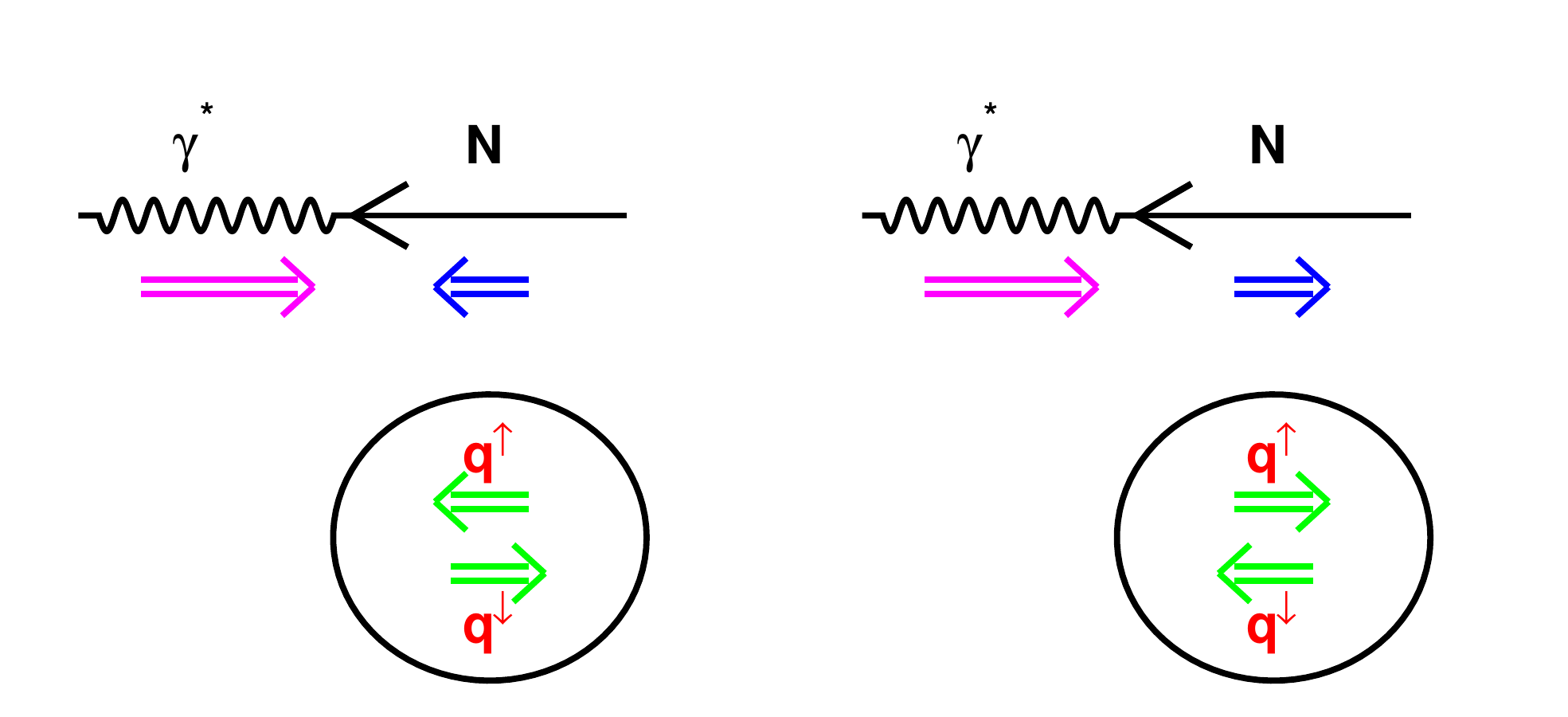}}
\caption{Accessing the helicity distributions in polarized deep inelastic
  scattering. 
Double arrows indicate the spin direction.\label{spineff}}
\end{figure}
In the left diagram photon and nucleon spins are antiparallel.
In this case the photon can only be absorbed by a quark 
having its spin aligned with the nucleon spin ($q^\uparrow$).
The absorption of the photon by the quark having its spin anti-aligned
with the nucleon spin would result in quark with $J_z=3/2$ in the final state
and is thus forbidden.
In a similar way, if photon and nucleon spin are aligned (Fig.~\ref{spineff}, right)
the photon can only be absorbed by a quark having its spin anti-aligned with
the nucleon spin ($q^\downarrow$).

The photon-nucleon cross section asymmetry
denoted by $A_1$ is thus given by 
\begin{equation}\label{a1}
  A_1(x) = \frac{\sigma^{\ua \da}_{\gamma^* N} - \sigma^{\ua \ua}_{\gamma^* N}}{\sigma^{\ua \da}_{\gamma^* N} + \sigma^{\ua \ua}_{\gamma^* N}}
   =  \frac{\sum_q e_q^2 (\Delta q(x) + \Delta \bar q(x))}{\sum_q e_q^2 (q(x) + \bar
     q(x))} \, ,
\end{equation}
$e_q$ being electric charge of the quark $q=u,d,s$.

Selecting specific hadronic final states, $h$, in semi-inclusive deep inelastic
scattering
\[
  \vec \ell  + \vec N \rightarrow \ell' + h + X
\]
allows to get more detailed information on the quark helicity distributions.
The corresponding photon-nucleon asymmetry reads 
\begin{equation}\label{a1_si}
  A_1^h(x,z) =  \frac{\sum_q e_q^2 \left( \Delta q(x) D_q^h(z) + \Delta \bar q(x) D_{\bar q}^h(z)\right)}
            {\sum_q e_q^2 \left(q(x) D_q^h(z) + \bar q(x) D_{\bar q}^h(z)\right)}.
\end{equation}
The fragmentation functions $D_q^h(z)$ describe the probability that
a quark $q$ fragments into a hadron $h$ carrying an energy fraction $z$
of the virtual photon energy $\nu$ in the target rest frame.
Semi-inclusive deep inelastic scattering allows to separate contributions
from quarks and anti-quarks because in general the corresponding fragmentation functions differ ($D_q^h \ne D_{\bar q}^h$).
Moreover, by selecting strange hadrons ($K^+$ and $K^-$) one can for example enhance
the contribution of the strange quarks. 

Another possibility to measure the quark helicity distributions
is polarized $pp$ scattering.
Single spin asymmetries of the process
\[
  \vec p + p \rightarrow W^{\pm} + \dots \rightarrow e^\pm + \dots 
\]
give access to quark helicity distributions
with the advantage that no knowledge on fragmentation functions is needed.
These single spin asymmetries are related to the helicity distributions
in the following way
\begin{eqnarray}
 A^{W^+}_L =  \frac{\Delta \bar d(x_1) u(x_2) - \Delta u(x_1) \bar d(x_2)} 
{u(x_1) \bar d(x_2) + \bar d(x_1) u(x_2) } \, ,\quad 
 A^{W^-}_L =  \frac{\Delta \bar u(x_1) d(x_2) - \Delta d(x_1) \bar u(x_2)} 
{d(x_1) \bar u(x_2) + \bar u(x_1) d(x_2) } \, .\nonumber
\end{eqnarray}

Accessing the gluon helicity distribution is more difficult.
It can for example be done by a next-to-leading order (NLO) analysis of spin
asymmetries where the leading order (LO) expression given in Eqs.~(\ref{a1})
and (\ref{a1_si})
are modified and receive contributions involving $\Delta g(x)$.
Another possibility is to look for hadronic final states tagging the
participation of gluons in the scattering process.
This can be done by selecting for example hadrons with large transverse momentum
with respect to the virtual photon axis or charmed mesons.
This is illustrated in Fig.~\ref{lo_pgf}. In a leading order process
where the photon is absorbed by one of the quarks, hadrons are essentially
produced along the axis of the virtual photon (Fig.~\ref{lo_pgf} left).
If a gluon participates in the partonic sub-process, hadrons 
can be produced at larger $p_T$ (Fig.~\ref{lo_pgf} right).
A particular clean tag of the gluon in the partonic subprocess 
is the detection of a charmed meson $D^0$ or $D^*$ in the final state,
because these are almost exclusively produced via the process 
$\gamma +g \rightarrow c + \bar c$.
\begin{figure}
\centering{\includegraphics[width=0.65\textwidth]{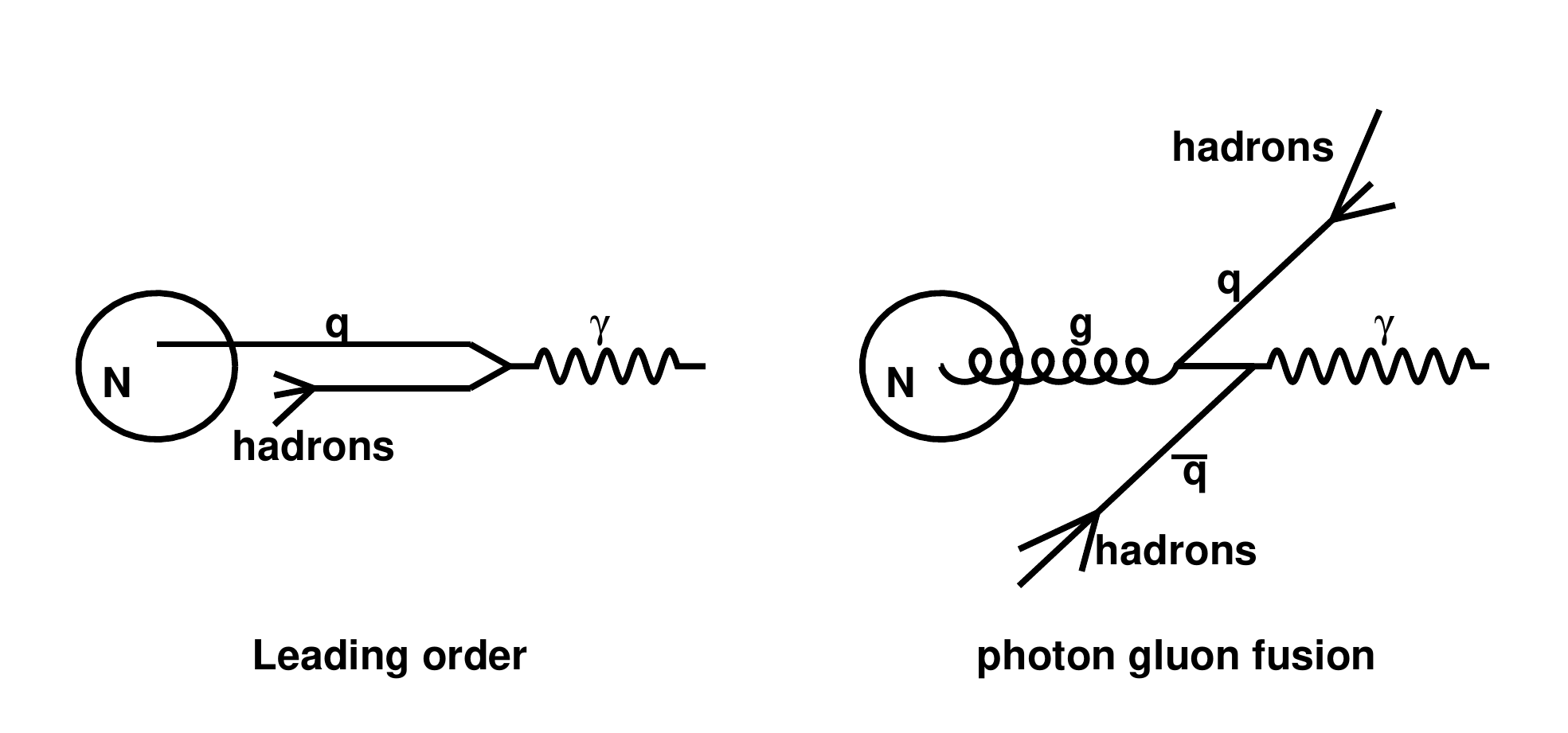}}
\caption{Leading Order (LO) and photon-gluon-fusion (PGF) process in deep inelastic scattering.
\label{lo_pgf}}
\end{figure}

In polarized $pp$ scattering double spin asymmetries for various final states
gives access to gluon helicity contribution.
These asymmetries 
depend on the product of parton helicity distributions $\Delta q^2$, $\Delta g^2$ and
$\Delta q \times \Delta g$.

\section{Results}\label{results}

\subsection{Quark Helicity distributions}
Figure~\ref{A1_p} shows the inclusive asymmetry $A_1^p$ obtained 
from several experiments in deep inelastic scattering of polarized electrons or muons
on polarized protons vs. the Bjorken variable $x$, which equals in LO the
parton momentum fraction, for a momentum transfer $Q^2>1$GeV$^2$.
Note that due to the different center of mass energies data points at a given $x$
may have different values of $Q^2$.
\begin{figure}[htb]
  \begin{center}
    \includegraphics[width=0.8\textwidth,clip,viewport=0 0 600 330 ]{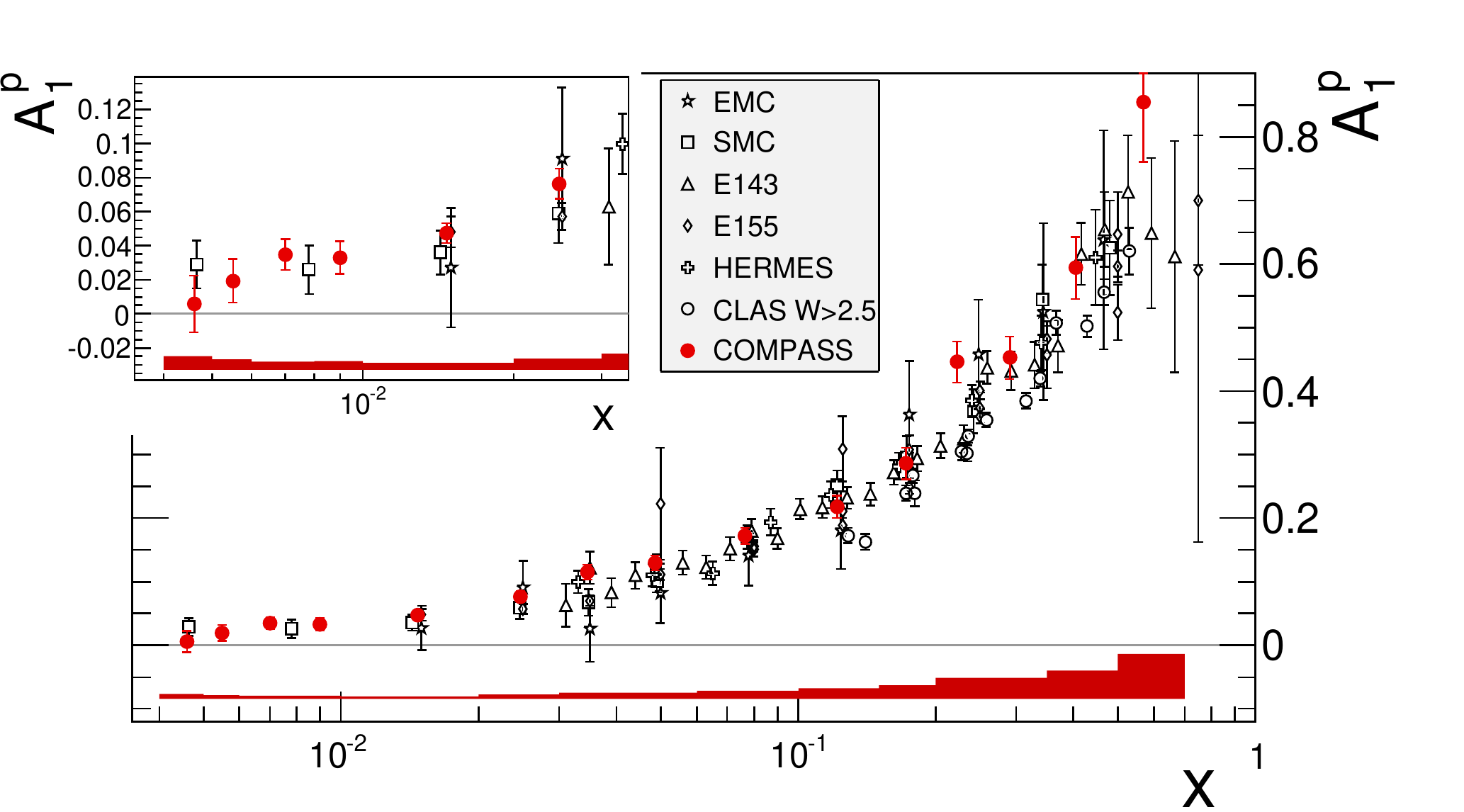}
    \caption{Results for the inclusive asymmetry $A_1^p$~\cite{compass_a1p}.}
    \label{A1_p}
  \end{center}
\end{figure}

Using these asymmetries together with the corresponding asymmetries for
neutron and deuteron and information from the neutron and hyperon decay
constants results in the values for the first moments of the quark distributions
given in Table~\ref{tab:dq}.
The most right column shows results from Lattice QCD calculations 
which are in remarkable agreement with the results obtained from experiment.
\begin{table}[tb]
    \caption{First moments of the polarized quark distributions at $Q^2=10$GeV$^2$.
 Note that the error for the result on the global analysis does not include 
an uncertainty for the unmeasured region 0<x<0.001.
    \label{tab:dq}}
  \begin{center}
\begin{tabular}{rcr|r}
\hline
& & global  analysis~\cite{dssv}  & lattice QCD~\cite{lqcd} \\
\hline \hline
 $\Delta \Sigma       $ &=&  $0.25 \pm 0.05$ & \\
 $\du + \dub$ &=&  $0.81 \pm 0.03$ & {$0.82\pm 0.04$}\\
 $\dd + \ddb$ &=& $-0.46 \pm 0.03$ & {$-0.41\pm 0.04$}\\
 $\ds + \dsb$ &=& $-0.11 \pm 0.06$ &  \\
\hline
\end{tabular}
  \end{center}
\end{table}

As discussed in Section~\ref{method} inclusive deep inelastic scattering 
gives only access to the sum quark and anti-quark contributions because $e_q^2 = e_{\bar q}^2$,
whereas semi-inclusive deep inelastic scattering allows to separate contribution
from quarks and anti-quarks. 
Figure~\ref{si_asy} shows semi-inclusive asymmetries measured on a proton target together
with the inclusive asymmetry from the COMPASS and HERMES experiment.
\begin{figure}
\includegraphics[width=\textwidth,viewport = 0 0 570 300,clip]{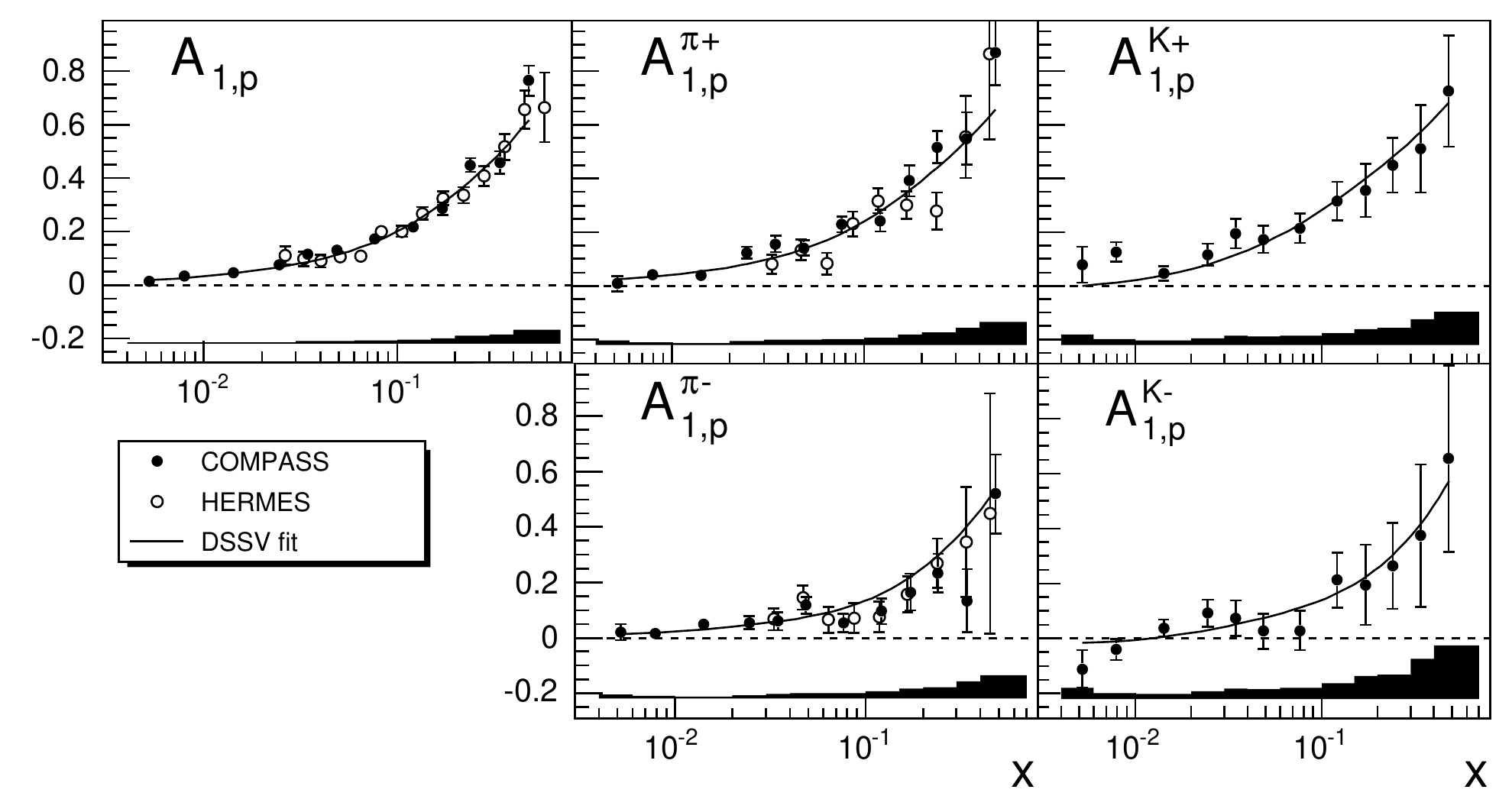}
 \caption{The inclusive and semi-inclusive asymmetry for $\pi^+$, $\pi^-$, $K^+$ and $K^-$
for the proton target vs. $x$ from the COMPASS and the HERMES experiment.\label{si_asy}}
\end{figure}

From these asymmetries together with asymmetries measured on the deuteron target the quark
helicity distributions are determined.
The analysis is done in leading order QCD where the measured asymmetries
are related to the helicity distribution via the following matrix equation:
\begin{eqnarray}
\vec A(x) &=& B\left(q(x),\int D_q^h(z) {\rm d}z\right)  
\, \Delta \vec q(x) \quad \mbox{with} \\
\vec A &=& (A_{1,p}, A_{1,p}^{\pi^+},A_{1,p}^{\pi^-},A_{1,p}^{K^+}, A_{1,p}^{K^-} ,
 A_{1,d},A_{1,d}^{\pi^+},A_{1,d}^{\pi^-},A_{1,d}^{K^+}, A_{1,d}^{K^-} ) \nonumber \\
\Delta \vec q&=&(\Delta u, \Delta d, \Delta s,\Delta \bar u, \Delta \bar d, \Delta \bar s) \nonumber
\end{eqnarray}
The matrix $B$ connecting the asymmetries with the helicity distributions
depends on the unpolarized quark distributions and the fragmentation functions.
Systematic errors originating form the choice of fragmentation functions
are discussed in the contribution of N.~Makke~\cite{nour}. 

Figure~\ref{fig:dq} shows the results assuming $\Delta s = \Delta \bar s$ published in Ref.~\cite{comp_si}.
Dropping this assumption, no difference was observed between $\ds$ and $\dsb$.
It only led to an increase in statistical error for the other helicity distributions.
$\du$ is positive and $\dd$ is negative mainly at large $x$.
The sea quark distributions are all close to zero over the whole measured range $0.004<x<0.3$.
For the strange quark one finds $\int_{0.004}^{0.3} \ds(x) {\rm d}x= -0.01 \pm 0.01
\pm 0.01$. The large negative value in Table.~\ref{tab:dq} comes mainly
from the unmeasured low $x$ region and is constrained from the neutron and hyperon decay
constants.
The truncated first moment $\int_{0.004}^{0.3} \Delta \bar u(x) - \Delta \bar d(x) \rm{d}x 
= 0.06 \pm 0.04(stat) \pm 0.02(syst)$
is slightly positive and disfavors models with $\Delta \bar u(x) - \Delta \bar d(x)<0$.
\begin{figure}
\centering{\includegraphics[width=0.8\textwidth, viewport= 0 0 570 390,clip]{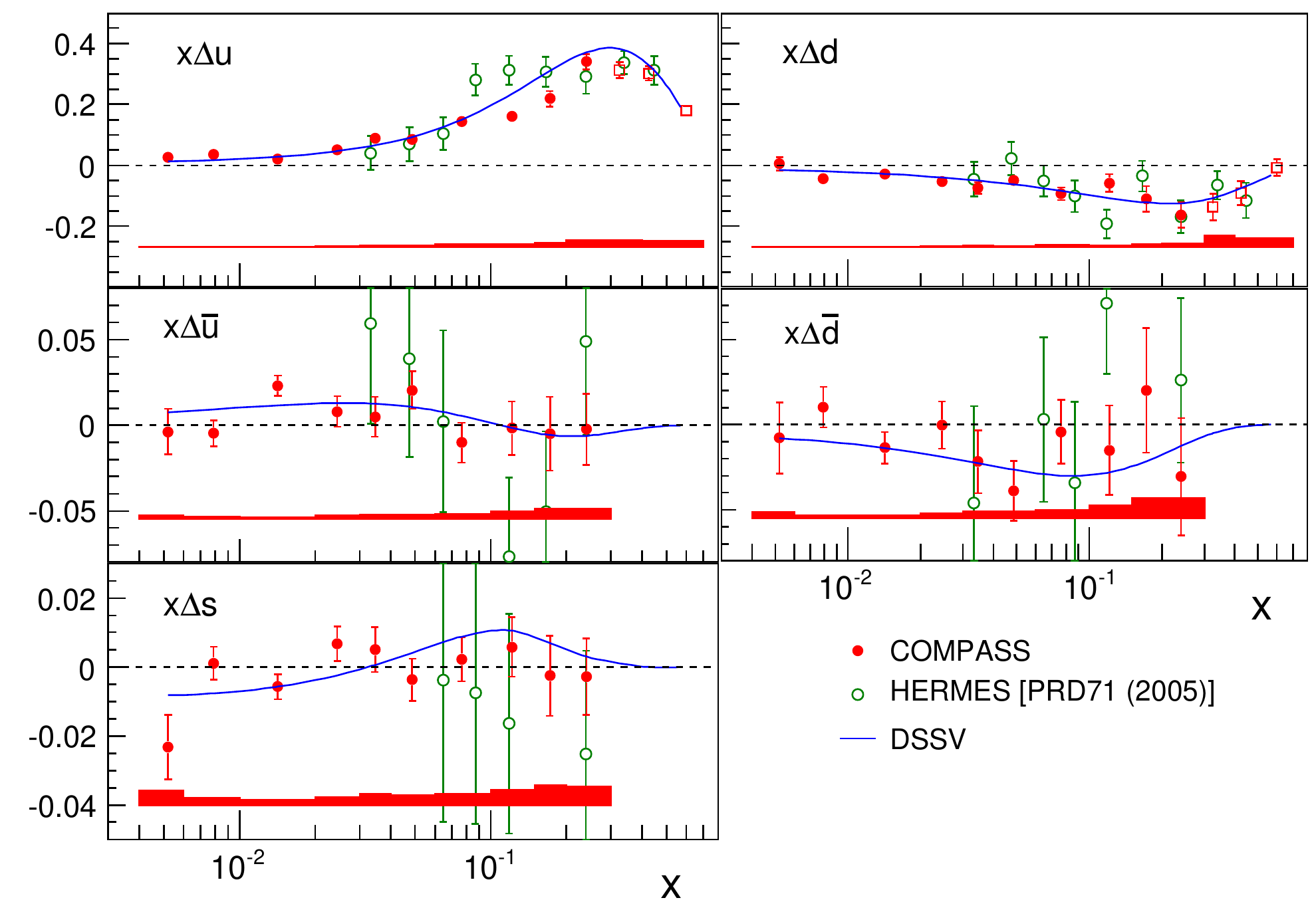}}
 \caption{The quark helicity distributions determined from inclusive and semi-inclusive asymmetries.\label{fig:dq}}
\end{figure}

First results on single spin asymmetries via the exchange of a $W^\pm$
measured by the PHENIX and STAR collaborations at $\sqrt{s}=500$ GeV at RHIC/BNL are shown in Fig.~\ref{fig:dq_rhic}.
The measured asymmetries have the expected sign, as can be seen by comparing
them to the curves in the Figure which result from various global analyses.
To compete with the results from deep inelastic scattering more data are needed.
\begin{figure}[h!]
\includegraphics[width=0.6\textwidth, viewport= 0 0 380 260,clip]{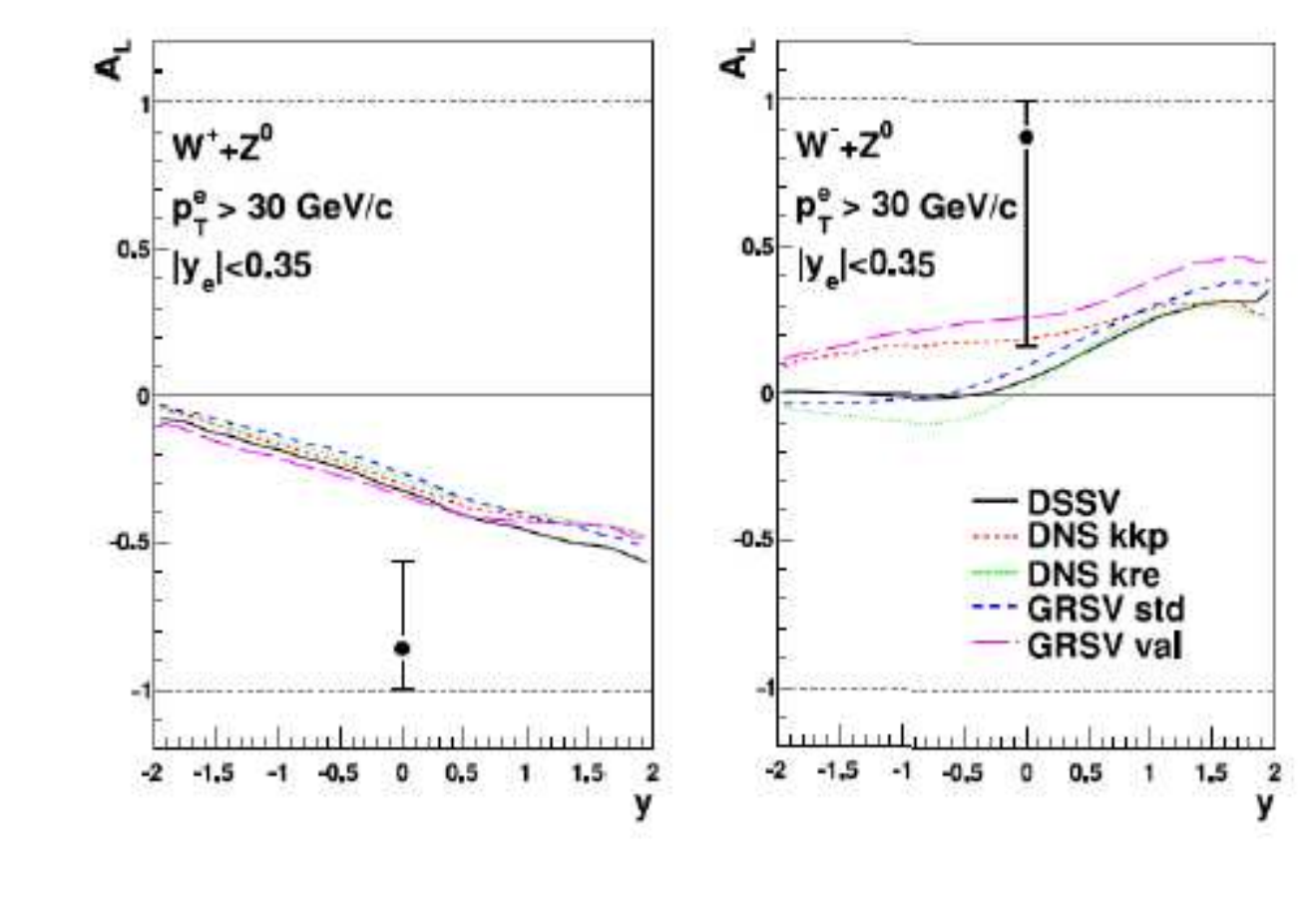}
\includegraphics[width=0.3\textwidth, viewport= 0 0 300 380,clip]{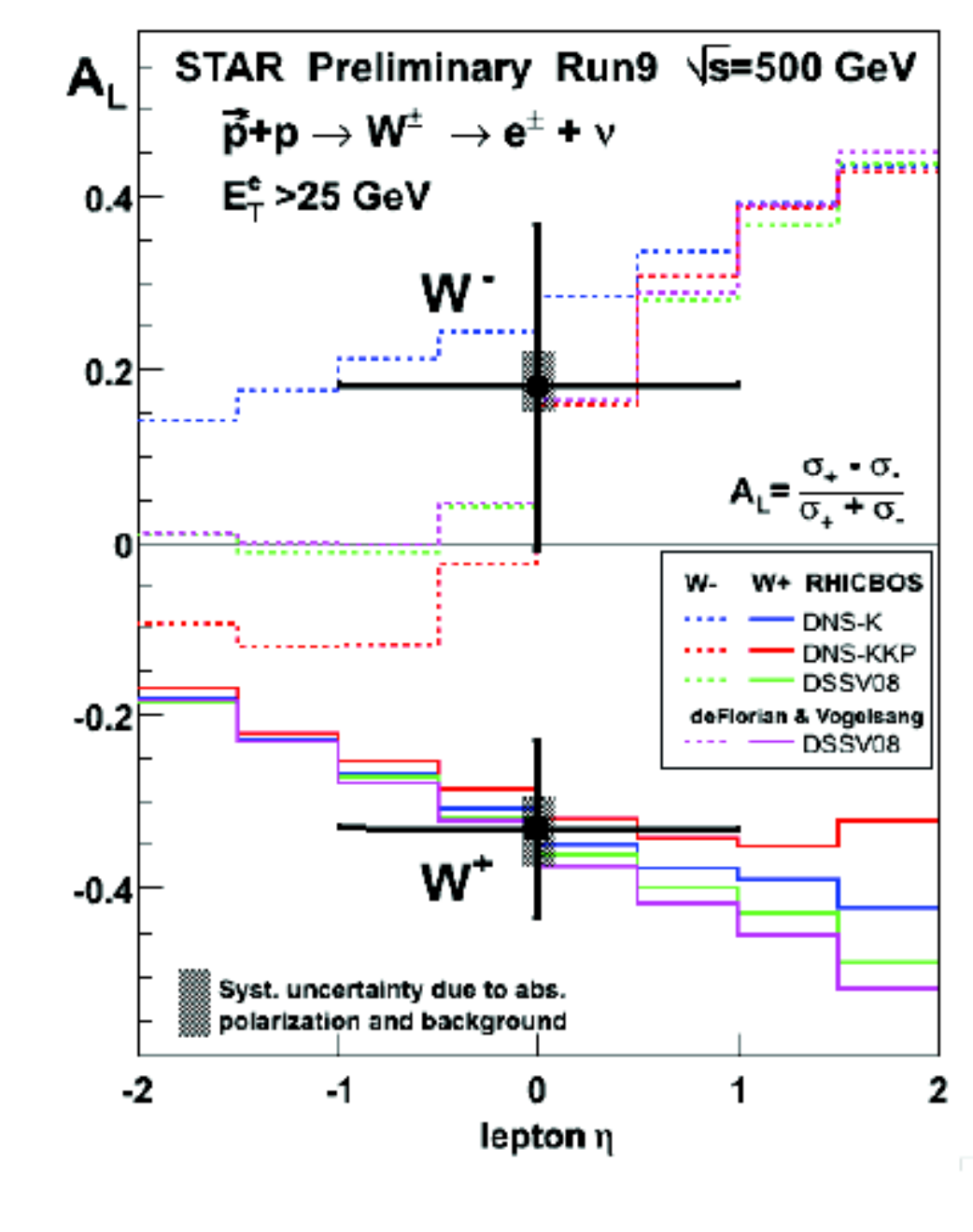}
\caption{The single spin asymmetry $A_L^{W^+}$ (left) and  $A_L^{W^-}$ (right) from
  the PHENIX~\cite{Haggerty:2010zz} experiment and the
  $A_L^{W^\pm}$ (right) from the STAR~\cite{star_W} experiment
as a function of the lepton rapidity compared to results from global analyses
of deep inelastic scattering data. 
\label{fig:dq_rhic}}
\end{figure}

\subsection{Gluon helicity distribution}
Figure~\ref{fig:dg_direct} (left) shows the result on $\Delta g/g$ from direct measurements 
from deep inelastic scattering using double spin asymmetries of high $p_T$
hadrons and open charm production. 
The measured asymmetries are directly related to the polarizations of gluons $\Delta g/g$:
\[
 A \propto \frac{\Delta g}{g} + A_{\mbox{bgd}} \, .
\] 
All results indicate that $\Delta g/g$ is small in the measured region $x_g \approx 0.1$
certainly excluding large values $\dG = 2-3$.
Here the data where analyzed in LO QCD~\footnote{The LO process in this
  context is the lowest order diagram involving a gluon, thus the PGF-diagram
  shown in Fig.~\ref{lo_pgf} (right)}.
For the open charm data also a NLO analysis is available. The result is shown 
in Fig.~\ref{fig:dg_direct} (right) together with results of various 
global analyses. 
Note that the vertical error bars indicate statistical
and systematic error, whereas the horizontal error bars indicate the 
$x_g$ range covered by the data. 
\begin{figure}
\includegraphics[width=0.5\textwidth, viewport = 0 0 550 390, clip]{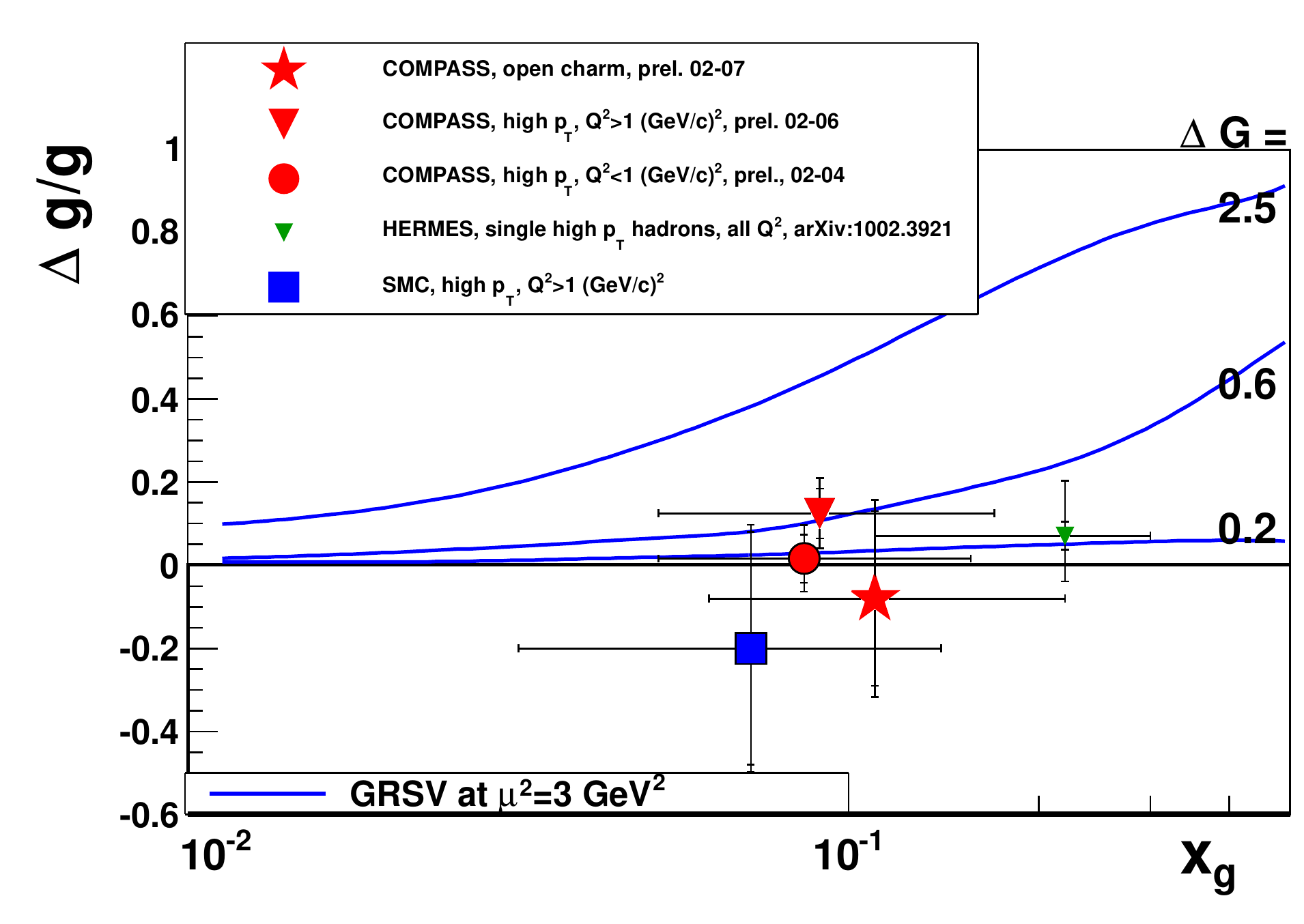}
\includegraphics[width=0.5\textwidth,viewport = 0 0 570 370, clip]{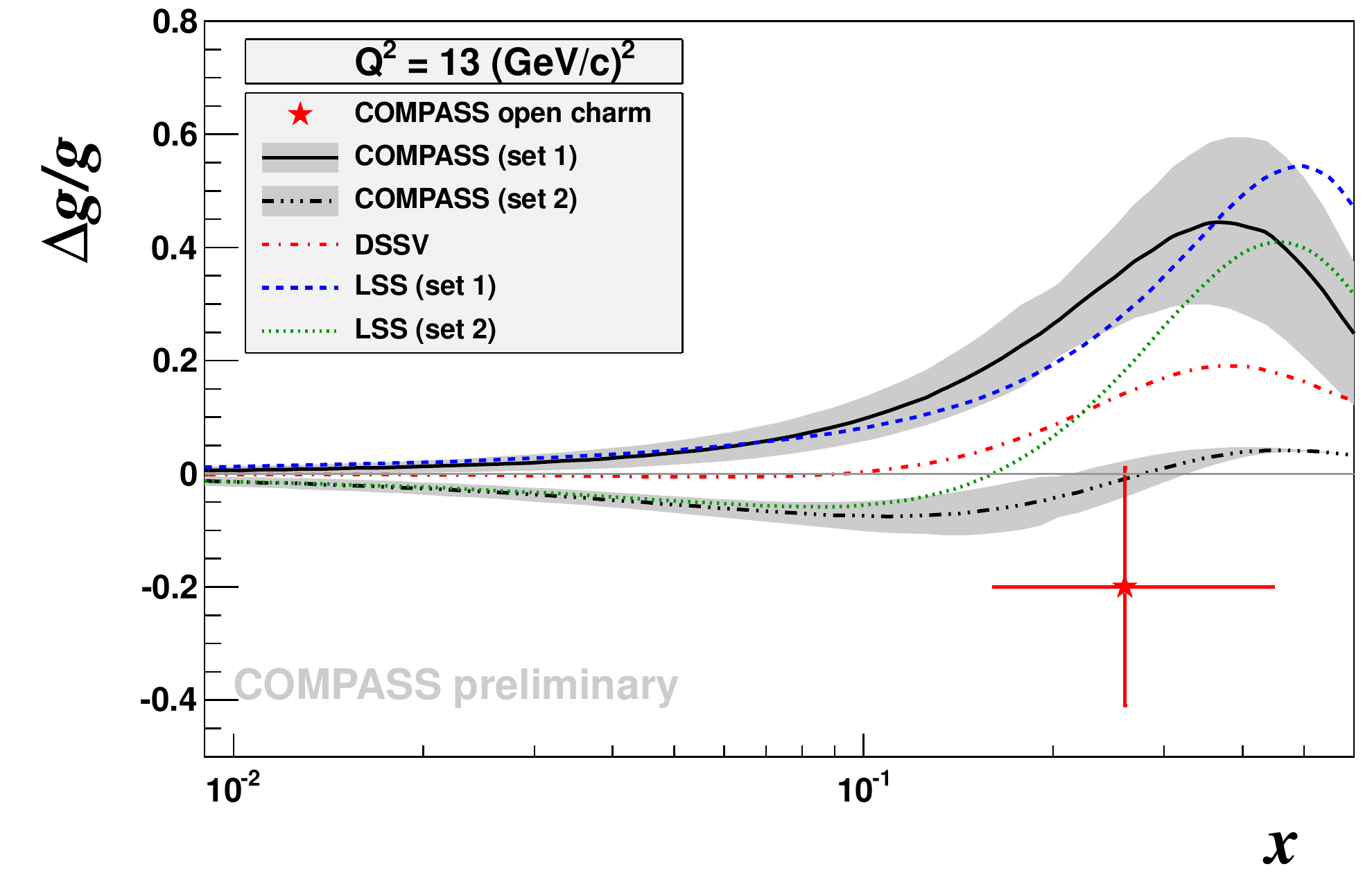}
\caption{Left: Results from direct measurements of $\Delta g/g$ using a
LO QCD analysis. The three blue lines are parameterization corresponding to three
different first moments of $\dG$. \\
Right: NLO result from the COMPASS open charm data together
with three analyses: COMPASS: global analysis of inclusive asymmetries
and the open charm measurement, LSS:  global analysis of inclusive and
semi-inclusive asymmetries~\cite{lss}, DSSV: global analysis discussed
below~\cite{dssv}.\label{fig:dg_direct}
}
\end{figure}

Double spin asymmetries in $pp$ scattering from $\pi^0$ production (PHENIX) and jet production (STAR) are shown in Fig.~\ref{fig:dg_rhic}.
Comparing these asymmetries measured as a function of the transverse momentum $p_T$
with parameterizations gives information on $\Delta g(x)$.
More results on $\Delta g(x)$ from the $pp$ scattering at the RHIC experiments are discussed in the contribution
of E. Aschenauer~\cite{elke}. 
\begin{figure}
\begin{minipage}[c]{0.48\textwidth}
\vspace{0pt}
\includegraphics[width=0.75\textwidth]{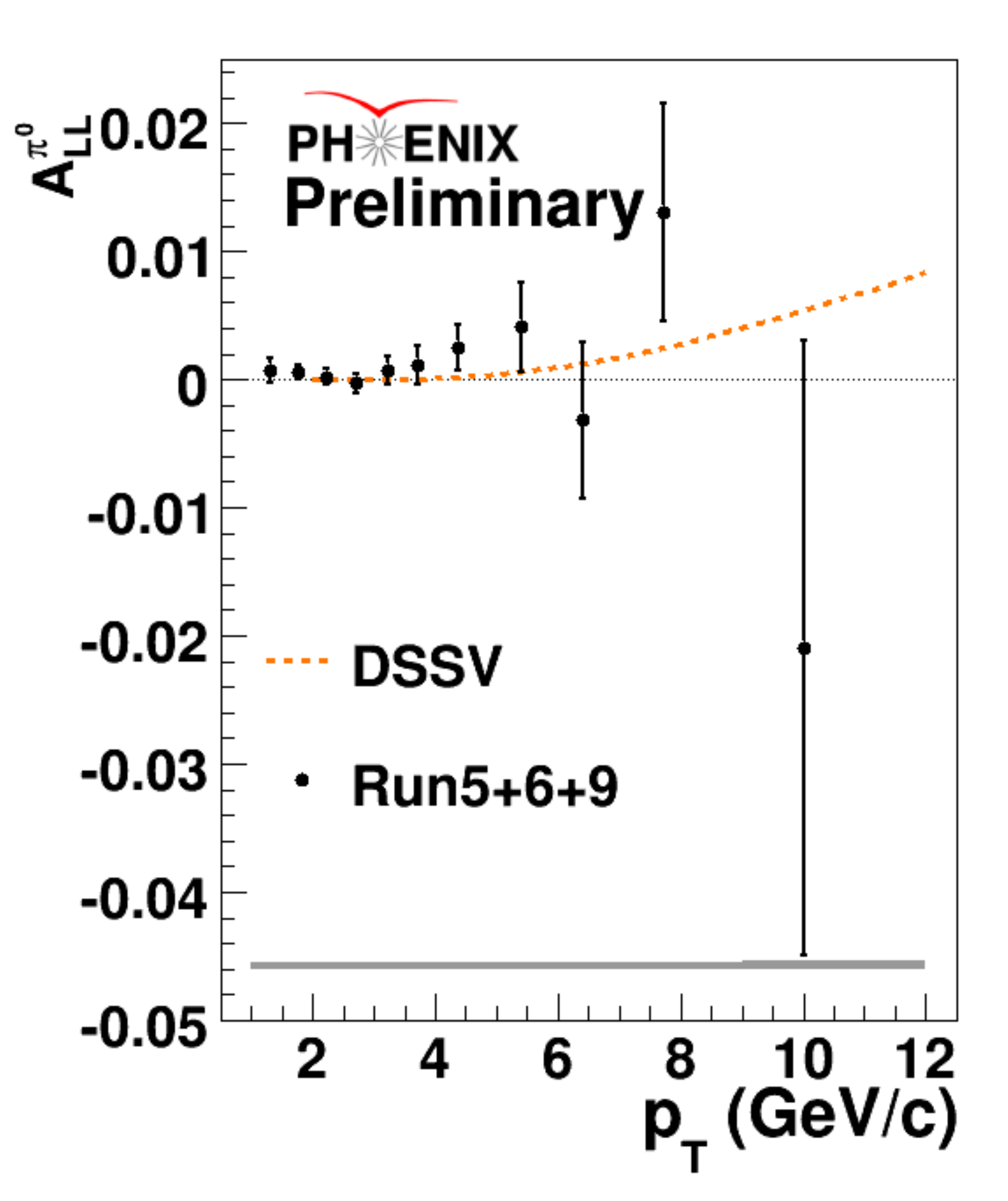}
\end{minipage}
\begin{minipage}[c]{0.48\textwidth}
\vspace{0pt}
\includegraphics[width=\textwidth]{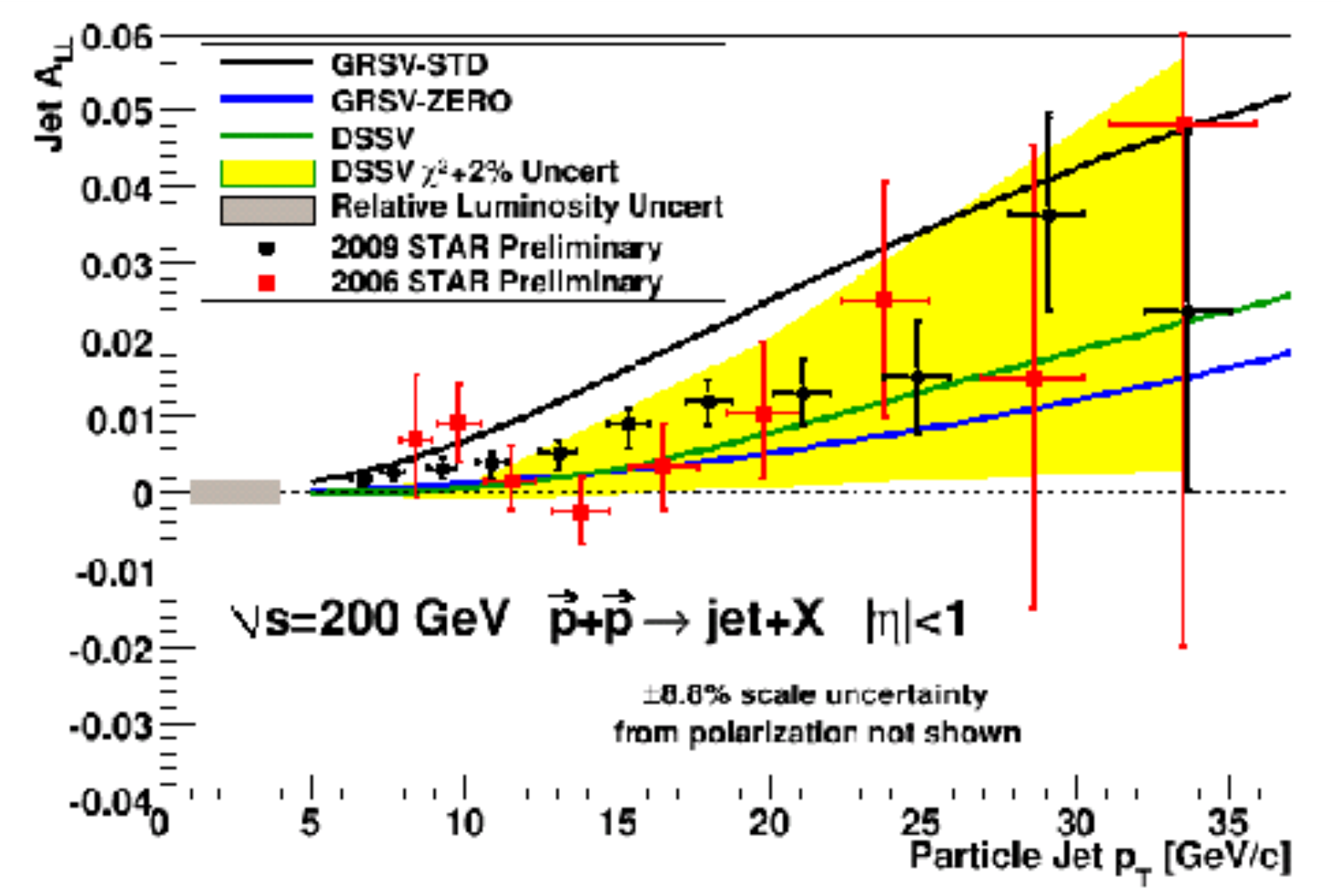}%
\end{minipage}
\caption{
Double spin asymmetries in $pp$ scattering from $\pi^0$ production
(PHENIX)
and jet production (STAR).
\label{fig:dg_rhic}}
\end{figure}

Results from a global NLO QCD analysis 
including inclusive, semi-inclusive asymmetries as well as asymmetries 
from $pp$ scattering are shown in Fig.~\ref{fig:dq_dssv}~\cite{dssv}.
Note that the newest COMPASS semi-inclusive results and most recent
RHIC results are not yet included in this analysis,
but they will not change the overall picture~\cite{marco_dis}.
\begin{figure}
\centering{\includegraphics[width=0.78\textwidth,viewport=0 0 520 407]{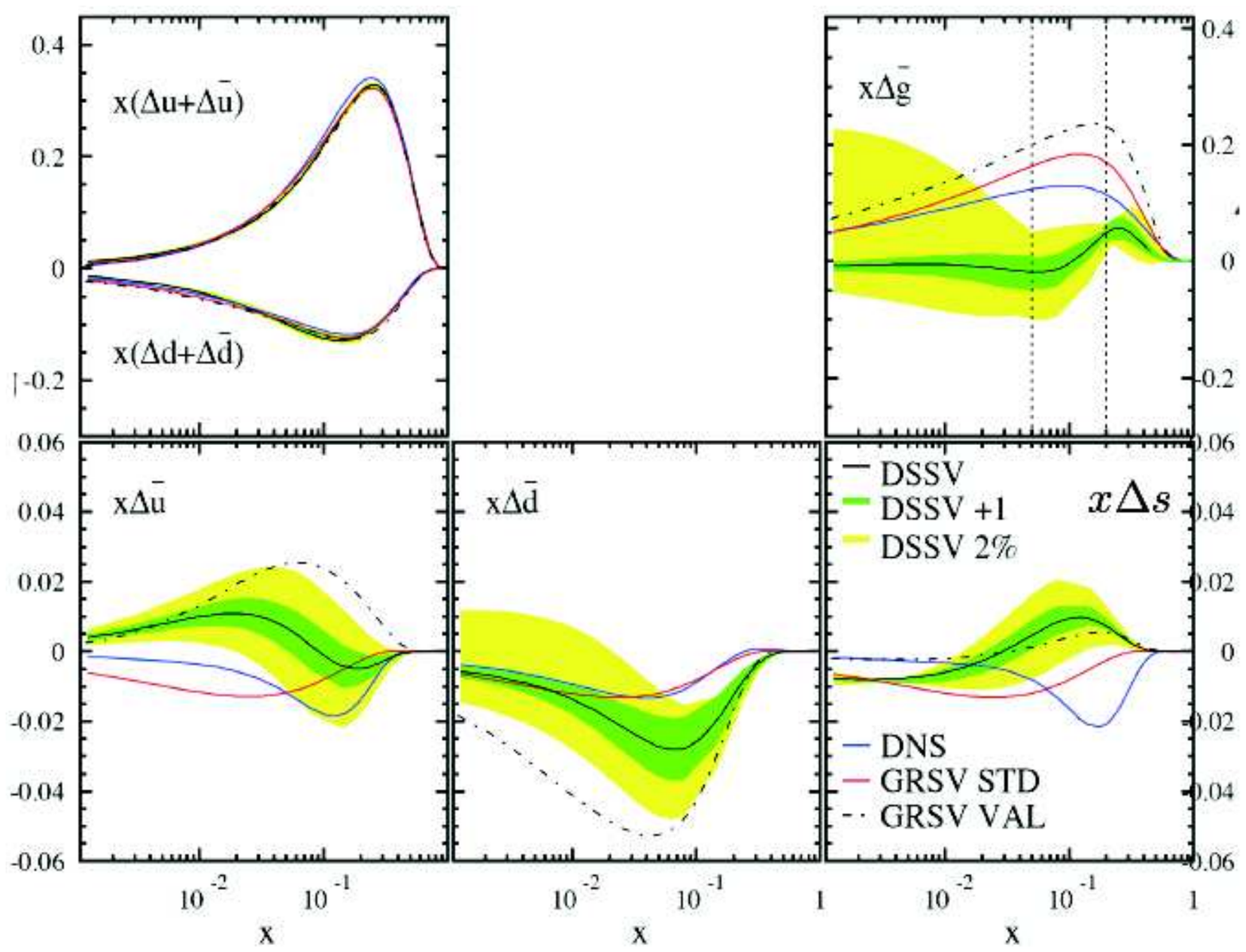}}
 \caption{Results on helicity distributions from a global analysis~\cite{dssv}.\label{fig:dq_dssv}}
\end{figure}

The results on $\du + \dub$ and $\dd + \ddb$ are driven by the inclusive asymmetries,
whereas the sea quark contributions are mainly determined by the semi-inclusive asymmetries. 
The largest influence on $\Delta g(x)$ comes from the $pp$ data at RHIC.
For the truncated first moment one finds
\[
 \int_{0.001}^{1} \Delta g(x) \rm{d}x = 0.013^{+0.702}_{-0.314}  \, .
\]

Note that the direct measurements from deep inelastic scattering are not yet included 
in this global analysis because for some channels a NLO QCD description is not yet available. Their inclusion will in the near future
further constrain $\Delta g(x)$.

\section{Future experimental programs}\label{future}
Table~\ref{tab:future} gives an overview over some important parameter of past, present and future experiments
in polarized deep inelastic scattering.
In the near future experiments at JLAB, CERN, and BNL will continue to take
data. Recently a new COMPASS proposal was accepted to study
so called Generalized Parton Distributions in Deep Inelastic Virtual Compton
Scattering (DVCS) and Hard Exclusive Meson Production (HEMP) as well
as Transverse Momentum Dependent distributions (TMDs) in Drell-Yan processes~\cite{eva,compassII}.
Similar programs are foreseen at JLAB and BNL.  

Table~\ref{tab:future} lists also projects for the far future.
The electron ion collider (EIC) projects in the US are discussed in the contribution of J.~Lee\cite{lee}.
The project discussed in Europe at GSI, a 3~GeV electron beam colliding with a 15~GeV
proton beam
has a center of mass energy and a luminosity comparable to the one of
the running COMPASS experiment.
The last line of the table gives the product squared of target, beam polarization and the target dilution factor.
This quantity is of particular interest for double polarization experiments.
Multiplied with the luminosity it gives an effective luminosity
which is proportional to the figure of merit (= inverse squared of the
statistical error) of the measured asymmetry.
In this quantity one would gain a factor $0.41/0.026 \approx 16$.
For the collider a polarization of 80\% for both beams is assumed.
In addition to this factor a large gain in hadron reconstruction is expected.
In the case of a collider one does not suffer from hadron interactions
in a solid state target. 
Such a collider, even at center of mass energies of already running experiments,
would thus offer great opportunities for studies of the spin structure of the nucleon.

\begin{table}[tb]
\caption{Parameters of experiments to study deep inelastic scattering.\label{tab:future}}
  \begin{center}
\begin{tabular}{|l|c|c|c|c|c|}
\hline
 Experiment   & JLab & HERMES & { ENC} & COMPASS & {EIC}\\
              & (12 GeV)             & @DESY        & @FAIR/GSI & @CERN  & @BNL/JLab\\
\hline
 $s$/GeV$^2$ & 23           &  50    & 180        &    300  & 10000\\
\hline
$x_{bj, min} = \frac{1 \mbox{GeV}^2}{y s}$    &     $5 \cdot 10^{-2}$
& $2 \cdot 10^{-2}$   &  $6 \cdot 10^{-3}$ &     $4 \cdot 10^{-3}$        &    $10^{-4}$              \\
for $y=0.9$  & & & & & \\
and $Q^2>1$GeV$^2$ & & & & & \\

\hline
${\cal L}$/(1/cm$^2$/s) &  $ \approx  10^{38}$   &
$\approx  10^{32}$    &{$ \approx 10^{32-33} $}       &   {$\approx  10^{32}$} & $\approx  10^{33-34}$\\
\hline
$(P_T P_B f)^2$ &  0.026     &      0.16                &  {0.41}                &  {0.026}         & 0.24 \\
\hline
\end{tabular}
\end{center}
\end{table}

\section{Summary \& Outlook}
It is well established since several years that the helicity contribution
of quarks to the nucleon spin is only about $\dS = 25\%$ in contrast to
a value of about 60\% expected in relativistic quark models.

New data, mainly from semi-inclusive double spin asymmetries from the COMPASS experiment
allowed to determine the contributions of the different quark flavors as a function
of the Bjorken variable $x$ to $\dS$.
More data will decide whether the difference in
the contribution of the light sea quarks $\dub$ and $\ddb$ is only 
a statistical deviation.
The RHIC experiment published first data on single spin asymmetries giving
a direct access to the helicity distributions.

Concerning the gluon helicity distribution, the results indicate
that the gluon distribution is small, but it should be kept in mind
the error on the first moment is still on the order of 1/2,
i.e. the gluon could still account for 100\% of the nucleon spin.

New experimental programs at JLAB, CERN and BNL will continue in the near future
measurements of the nucleon structure, where the interest will be enlarged  
to measurements of 
Generalized Parton Distributions (GPDs) and Transverse Momentum Dependent
distributions (TMDs)  
giving also access to orbital angular momentum contributions.
In the far future new facilities, like a polarized electron nucleon collider would offer great
potential to study the spin structure of the nucleon at a much deeper level.


%

}  


\end{document}